\documentclass[prd,tightenlines,floatfix,showpacs,preprintnumbers,nofootinbib,eqsecnum]{revtex4}
 \usepackage[dvips,final]{graphicx}
  \usepackage{amssymb}
   \usepackage{amsmath} 
    \usepackage{amsfonts}
     \usepackage{epsfig}
     \usepackage{color}
      \usepackage{bm} 
        \usepackage{xcolor}
        
\usepackage{booktabs}
\usepackage{multirow}
\usepackage{slashed}
\usepackage{comment}

\def\doublespace{\def\baselinestretch{1.6}\large\normalsize}
\def\normalspace{\def\baselinestretch{1.0}\normalsize}
\def\PSfig#1#2{\scalebox{#1}{\includegraphics{#2}}}

\def\Caption#1{
  \normalspace
  \vskip-1mm\caption{\sl#1}\vskip-1mm
  \doublespace
}

\def\BA{\begin{eqnarray}}
\def\BE{\begin{equation}}
\def\BF{\begin{figure}[htb]}
\def\BT{\begin{table}[htb]}
\def\EA{\end{eqnarray}}
\def\EE{\end{equation}}
\def\EF{\end{figure}}
\def\ET{\end{table}}

\def\ra{\rangle}

\def\TeV{\,\mbox{TeV}}
\def\GeV{\,\mbox{GeV}}

\def\Jpsi{J\!/\!\psi}
\def\psip{\psi^{\,\prime}}

\def\Y{\Upsilon}
\def\Yp{\Upsilon^{\,\prime}}

\def\sqq{\sigma_{Q\bar Q}}

\def\aem{\alpha_{em}}

\def\lsim{\mathrel{\rlap{\lower4pt\hbox{\hskip1pt$\sim$}}
     \raise1pt\hbox{$<$}}}         
 \def\gsim{\mathrel{\rlap{\lower4pt\hbox{\hskip1pt$\sim$}}
     \raise1pt\hbox{$>$}}}         

\begin{document}

\title{
\vspace*{-2.0cm}
Ultra-peripheral nuclear collisions as a source of heavy quarkonia
%
}

\author{B. Z. Kopeliovich$^1$}
\email{boris.kopeliovich@usm.cl}

\author{M. Krelina$^{2,3}$}
\email{michal.krelina@cvut.cz}

\author{J. Nemchik$^{2,4}$}
\email{jan.nemcik@fjfi.cvut.cz}

\author{I. K. Potashnikova$^1$}
\email{irina.potashnikova@usm.cl}

\vspace*{0.5cm}

\affiliation{
\vspace*{0.2cm}
$^1$Departamento de F\'{\i}sica,
Universidad T\'ecnica Federico Santa Mar\'{\i}a,
Avenida Espa\~na 1680, Valpara\'iso, Chile}
\affiliation{$^2$
Czech Technical University in Prague, FNSPE, B\v rehov\'a 7, 11519
Prague, Czech Republic}
\affiliation{$^3$
Physikalisches Institut, University of Heidelberg, Im Neuenheimer Feld 226, 69120 Heidelberg, Germany}
\affiliation{$^4$
Institute of Experimental Physics SAS, Watsonova 47, 04001 Ko\v sice, Slovakia
}

\vspace*{2.0cm}
\date{\today}
\begin{abstract}
\vspace*{5mm}
Heavy quarkonium production in ultra-peripheral nuclear collisions (UPC)
is described within the QCD dipole formalism.
Realistic quarkonium wave functions in the $Q\bar Q$ rest frame are calculated by solving the Schr\"odinger equation with a subsequent  Lorentz boost to high energy. We rely on several realistic $Q\bar Q$ potentials, which allow to describe well the quarkonium masses and decay widths, as well as data on diffractive electroproduction of quarkonia on protons.
Nuclear effects are calculated with the phenomenological dipole cross sections fitted to DIS data. The higher twist quark shadowing related to the lowest $Q\bar Q$ Fock component of the photon, as well as the leading twist gluon shadowing, related to higher components containing gluons, are included.
The results for coherent and incoherent photoproduction of charmonia and bottomonia in UPC of heavy nuclei are in good accord with available data from the LHC. They
can also be verified in future experiments at electron-ion colliders. 
\end{abstract}


\maketitle

%
%
%
\vspace*{-0.25cm}
\section{Introduction}
\label{Sec:intro}
%
%
%

Ultra-peripheral collisions (UPC) of nuclei at high energies correspond to collisions with impact parameter much larger than the sum of nuclear radii. Correspondingly, the total cross section of UPC is much larger than that of collisions with nuclear overlap. Besides, UPC is the dominant source of heavy vector mesons. Collisions with double rapidity gaps can be interpreted as photon-Pomeron fusion into a heavy vector meson $V=\Jpsi(1S)$, $\Y(1S)$, or their radial excitations.
In impact parameter space the Pomeron is short range exchange, while the radius of photon exchange is infinitely large.

UPC provide a unique access to photon-nucleus interactions. Weizs\"acker-Williams photons, originated from one of the nuclei, can interact diffractively with another one. The mechanisms of photo-production of vector mesons on nuclei have been studied pretty well beyond the Glauber approximation, either in the multi-channel approach \cite{Hufner:1997jg}, or employing the color dipole model \cite{Kopeliovich:1991pu} .
Having a comprehensive understanding of photo-production mechanisms, one can switch to description of nuclear UPC, as was done in Ref.~\cite{Ivanov:2007ms}.

Within the popular light-front (LF) color dipole approach
\cite{Kopeliovich:1991pu,Kopeliovich:1993gk,Nemchik:1996pp,Kopeliovich:1993pw,Nemchik:1994fp,Nemchik:1994fq,Nemchik:1996cw,Hufner:2000jb,Nemchik:2000de,Nemchik:2000dd,Kopeliovich:2001xj,Kopeliovich:2007wx},
the effect, known as color transparency (CT), significantly affects heavy quarkonium photo-production off nuclei.
It is controlled by the {\it formation length} $l_f$, characterizing evolution of the $Q\bar Q$ pair, evolving from the small initial size $\sim1/m_Q$, where $m_Q$ is the heavy quark mass, up to a larger non-perturbative size of the quarkonium.
The corresponding expression for $l_f$
can be obtained in the nuclear rest frame from the condition that the relative phase shift between the two lowest levels, $V$ and $V^\prime$ becomes of the order of unity \cite{Kopeliovich:1991pu,Kopeliovich:2001xj}, 
%
 \BE
l_f = \frac{2\,k}
{M_{V^\prime}^2 - M_V^2}\,
\label{tf}
 \EE
%
where $k$ is the photon energy
and $M_V$ and $M_{V^\prime}$ are the quarkonium masses in $1S$ and $2S$ states, respectively.

Another important length scale, called {\it coherence length} (CL) \cite{Kopeliovich:1991pu,Kopeliovich:2001xj}, characterizes the phase shift between $Q\bar Q$ photo-production  amplitudes with different longitudinal coordinates of interaction.
It has the following form,
%
 \BE
l_c = \frac{2\,k}{M_V^2} ,
\label{lc}
 \EE
The amplitudes are coherent, provided that the length interval $\Delta l\ll l_c$.
The coherence length $l_c$ can also be interpreted as a lifetime (or path) of the $Q\bar Q$ fluctuation of the photon.

In our calculations of nuclear effects we rely, for the sake of simplicity, on the eikonal approximation for in-medium propagation of long-lived $Q\bar Q$ photon fluctuations, which is relevant at rapidity $y=0$.
However, at forward or backward rapidities at the Large Hadron Collider (LHC),
as well as at Relativistic Heavy Ion Collider (RHIC),
such an approximation cannot be applied anymore because the coherence length Eq.~(\ref{lc}) becomes too short at least for one of the colliding nuclei. Here, for the first time, we apply the rigorous path-integral technique
as is described in Sect.~\ref{fcor}.

Another source of nuclear suppression, called {\it gluon shadowing} (GS), in terms of parton model looks like reduction of gluon density in nuclei at small Bjorken $x$. In the infinite momentum frame of the nucleus this occurs due to longitudinal overlap and fusion of gluons, originated from different bound nucleons. This effect is difficult to evaluate, and usually extracted from the global fits to data. The result is unreliable and is known only for the density integrated over impact parameter.

The parton model description is not Lorentz invariant (only observables are), it depends on reference frame. What looks like gluon fusion in the infinite momentum frame of the nucleus, corresponds to usual Glauber-like shadowing of the photon fluctuations in the nuclear rest frame. Namely, gluon reduction corresponds to shadowing of higher Fock components of the photon, which contain one or more gluons (corrected for the effect for the lowest Fock state $|Q\bar Q\ra$) \cite{Kopeliovich:2022jwe}. The transverse size of such fluctuations depends on the hard scale ($m_Q$) logarithmically, so is the leading twist effect.

Notice that onset of gluon shadowing requires much smaller $x$ in comparison with the $Q\bar Q$ component. That happens due to a specifically short coherence length for higher Fock states, $l_c^G\ll l_c$. They differ by an order of magnitude \cite{Kopeliovich:2000ra,Kopeliovich:2022jwe}.
Since the transverse size of $Q\bar Q-G$ dipoles fluctuates during propagation through the nucleus,
even at very high energies of the LHC, one cannot rely on the "frozen" eikonal approximation,
$l_c^G\gg R_A$, where $R_A$ is the nuclear radius.
For this reason, our calculations of the GS effect are performed relying
on the Green function formalism \cite{Kopeliovich:1999am,Ivanov:2002kc,Kopeliovich:2022jwe}.

In the present paper we incorporate various improvements in theoretical description of quarkonium photo-production off nuclei, performed in Ref.~\cite{Kopeliovich:2022jwe}, and apply them also to UPC at RHIC and the LHC. We include  
the higher- and leading-twist shadowing corrections corresponding to the $|Q\bar Q\ra$ and
$|Q\bar QG\ra$ Fock state of the photon, respectively. 
The multi-gluon Fock components have too short coherence length to generate a significant shadowing effects even at very high energies of the LHC.
Moreover, as is discussed in Ref.~\cite{Kopeliovich:2022jwe} (see also Refs.~\cite{Krelina:2018hmt,Cepila:2019skb,Krelina:2019egg,Krelina:2020bxt}), we ignore the frequently used unjustified model of the photon-like $V\to Q\bar Q$ transition,
which would lead to an exaggerated weight of the $D$-wave in the rest frame quarkonium wave function.

The paper is organized as follows. We present expressions for calculations of differential cross sections $d\sigma/dy$ corresponding to coherent (elastic) and incoherent (quasi-elastic) heavy quarkonium production in UPC  in Sections~\ref{sub-coh} and \ref{sub-inc}, respectively. 
Incorporation of a small real part of the production amplitude, as well as spin rotation effects
is explained in Sec.~\ref{sub-sr}.
In the following Section~\ref{fcor} we discuss the two main effects, which influence on 
nuclear effects in UPC: 
(i) the corrections for a finite coherence length and 
(ii) gluon shadowing. 
The former effect is calculated for the first time within the LF dipole approach based on the Green function formalism, leading to the results, that are substantially different from the standard vector dominance model (VDM). 
The following Sec.~\ref{res} is devoted to comparison of model predictions with available data and to the analysis of particular nuclear effects in coherent and incoherent quarkonium production in UPC. 
Finally, the last Sec.~\ref{conclusions} contains a summary and concluding remarks.

%
%
%
\section{Quarkonium production cross section in 
ultra-peripheral collisions}
\label{Sec:cross-sec}
%
%
%

The large charge $Z$ of colliding heavy nuclei gives rise to strong electromagnetic
fields: in a heavy-ion UPC, the photon field of one nucleus can 
produce a photo-nuclear reaction in the other. Then the cross section for photo-production of a vector meson $V$ by the Weizs\"acker-Williams photons
can be written in the rest
frame of the target nucleus $A$ as follows \cite{Bertulani:2005ru}:
%
\BE
  k\frac{d\sigma}{dk} = \int\,d^2\tau \int\,d^2b \,\,
  n(k,\vec b-\vec\tau,y)\, 
  \frac{d^2
  \sigma_A(s,b)}{d^2b}
  ~~ + ~~
  \Bigl \{ y\rightarrow -y \Bigr \}
  \,,
\label{cs-upc}
\EE
%
where the rapidity variable $y = \ln \bigl [s / (M_V \sqrt{s_N}~) \bigr] \approx \ln\bigl [(2 k M + M^2) / (M_V \sqrt{s_N}~)\bigr ]$.

The formula (\ref{cs-upc}) is derived in the one-photon-exchange approximation. Here
the variable $\vec\tau$ is the relative impact parameter of a nuclear collision, and $\vec{b}$ is the impact parameter of the photon-nucleon collision
relative to the center of one of the nuclei. 
Particularly, 
the collision of identical nuclei with the nuclear radius $R_A$ in UPC
leads to a condition that the impact parameter $\tau > 2 R_A$ \cite{Bertulani:2005ru}.

The variable
$n(k,\vec b)$ in Eq.~(\ref{cs-upc}) represents the photon flux induced by the
projectile nucleus with Lorenz factor $\gamma$ and has the following form, 
%
\BE
  n(k,\vec b) = \frac{\aem Z^2 k^2}{\pi^2\gamma^2}
  \Biggl [
  K_1^2\left(\frac{bk}{\gamma}\right)
  +
  \frac{1}{\gamma^2} K_0^2\left(\frac{bk}{\gamma}\right)
  \Biggr ]
  \,,
  \label{flux}
\EE
%
where $\aem= 1/137.036$ is the fine-structure constant,
$K_{0,1}$ are the modified Bessel functions of the second kind and the Lorentz factor $\gamma = 2 \gamma_{col}^2 - 1$ with $\gamma_{col} = \sqrt{s_N}/2 M$.
The first and the second
term in Eq.~(\ref{flux}) corresponds to the flux of photons transversely and 
longitudinally polarized to the ion direction, respectively. The former photon flux
dominates in ultra-relativistic collisions with $\gamma\gg 1$. Consequently, in heavy-ion UPC
at RHIC and the LHC one can safely neglect the second term in Eq.~(\ref{flux}) treating the photons as almost real due to very small virtuality, $-q^2 = Q^2 < 1/R_A^2$.

%
%
%
\subsection{Coherent production of quarkonia}
\label{sub-coh}
%
%
%

To calculate the cross sections for coherent ($coh$) 
quarkonium production, $\gamma A\to V\!A$, we use the light-front dipole approach
\cite{Kopeliovich:1991pu}, which has been applied to describe $\Jpsi$ photo-production
off nucleons \cite{Hufner:2000jb,Krelina:2018hmt,Cepila:2019skb} and nuclei 
\cite{Ivanov:2002kc,Nemchik:2002ug}. In this approach, 
assuming sufficiently large photon energies, corresponding to most of the kinematic regions studied in the present paper when the CL (\ref{lc}) 
$l_c\gg R_A$,
the nuclear cross section takes a simple asymptotic form,

%
 \BA
 \frac{d^2\sigma_{A}^{coh}(s,b)}{d^2b} \Biggr|_{l_c \gg R_A} 
&=&
\Biggl |
\int d^2r\int_0^1 d\alpha\,
\Psi_{V}^{*}(\vec r,\alpha)\,
\Biggl (1 -
\exp\left[-\frac{1}{2}\,\sqq(r,s)\,T_A(b)\right]
\Biggr )
\Psi_{\gamma}(\vec r,\alpha)
\Biggr |^2
\nonumber\\
&\equiv&
\Biggl |
\int d^2r\int_0^1 d\alpha\,
\Sigma_{A}^{coh}(r,\alpha,s,b)\,
\Biggr |^2\,.
\label{coh-lcl}
 \EA
%
Here we rely on the optical approximation, assuming the elastic dipole amplitude pure imaginary, which is rather accurate for heavy nuclei.
Expression (\ref{coh-lcl}) is frequently called ``frozen'' approximation, assuming that the transverse separation of the $|Q\bar Q\ra$ Fock state of the photon does not change
during propagation through a nuclear medium. It represents the higher twist shadowing correction since the $Q-\bar Q$ transverse separation diminishes as $1/m_Q$.

In Eq.~(\ref{coh-lcl}) 
$T_A(b) = \int_{-\infty}^{\infty} dz\,\rho_A(b,z)$ is the nuclear thickness function normalized as $\int d^2 b\,T_A(b) = A$, 
where $\rho_A(b,z)$ is the nuclear density function, for which we employ the realistic Wood-Saxon form with parameters taken from \cite{DeJager:1987qc};
$\Psi_V(r,\alpha)$ is the LF wave function for heavy quarkonium;
$\Psi_{\gamma}(r,\alpha)$ is the LF distribution or the wave function 
of the $Q\bar Q$ Fock component of the quasi-real (transversely polarized) photon,
where the $Q\bar Q$ fluctuation (dipole) has the transverse size $\vec{r}$ and the variable $\alpha = p_Q^+/p_{\gamma}^+$ is the boost-invariant fraction of the photon momentum carried by a heavy quark (or antiquark).

The universal dipole-nucleon total cross section $\sqq(r,s)$ depends on transverse dipole separation $r$ and c.m. energy squared $s = M_V\,\sqrt{s_N}\,\exp[y]$. Energy dependence of the dipole cross section can be alternatively included also via variable $x = M_V^2/s = M_V\,\exp[-y]/\sqrt{s_N}$.

Notice that the coherent cross section, Eq.~(\ref{coh-lcl}), is different from the usual Glauber expression \cite{Bauer:1977iq} due to presence of the dipole cross section \cite{Kopeliovich:1981pz}. It effectively includes the 
Gribov inelastic shadowing corrections \cite{Gribov:1968jf,Kopeliovich:2016jjx} in all orders for the $Q\bar Q$ Fock component of the photon.

%
%
%
\subsection{Incoherent production of quarkonia}
\label{sub-inc}
%
%
%

Besides ``elastic'' coherent photo-production $\gamma A\to V A$, where the nucleus remains intact, the vector meson can be produced in a quasi-elastic process $\gamma A\to V A^*$, where the nucleus is excited and decays to fragments. Important is that additional meson production is excluded. In this case, one can sum over different products of  nuclear excitation and employ the conditions of completeness. Of course, one channel of elastic photo-production must be subtracted. It is instructive to see the result within the Glauber approximation \cite{Hufner:1996dr},
%
 \BA
 \!\!\!\!\!
 \frac{d^2\sigma_A^{inc}(s,b)}{d^2b}
\Biggr|_{l_c \gg R_A}^{Gl} 
\propto
\exp\left[-\sigma^{VN}_{in}\,T_A(b)\right] -
\exp\left[-\sigma^{VN}_{tot}\,T_A(b)\right]
=
\exp\left[-\sigma^{VN}_{tot}\,T_A(b)\right]
\Bigl\{\exp\left[-\sigma^{VN}_{el}\,T_A(b)\right]-1 \Bigr\}\,.
\label{glauber1}
\EA
%
Here the inelastic $V-N$ cross section $\sigma^{VN}_{in}=
\sigma^{VN}_{tot}-\sigma^{VN}_{el}$, where the elastic cross section
%
\BA 
\sigma^{VN}_{el} \approx\frac{(\sigma^{VN}_{tot})^2}{16\,\pi\,B^{VN}}
\label{glauber2}
\EA
%
and $B^{VN}$ is the slope of the differential elastic $V-N$ cross section.

The cross section of incoherent ($inc$) photo-production has the form, analogous to (\ref{glauber1}), but with additional integrations over the dipole size (see derivation in Sect.~VII of Ref.~\cite{Kopeliovich:2005us}),
%
 \BA
\frac{d^2\sigma_A^{inc}(s,b)}{d^2b}
\Bigr|_{l_c \gg R_A} 
&= &
\int d^2r_1\int_0^1 d\alpha_1\,
\Psi_{V}^{*}(\vec r_1,\alpha_1)\,\Psi_{\gamma}(\vec r_1,\alpha_1)\,\exp\left[-\frac{1}{2} \sqq(r_1,s) T_A(b)\right]
\nonumber\\
&\times &
\int d^2r_2\int_0^1 d\alpha_2\,
\Psi_{V}^{*}(\vec r_2,\alpha_2)\,\Psi_{\gamma}(\vec r_2,\alpha_2)\,\exp\left[-\frac{1}{2} \sqq(r_2,s)\, T_A(b)\right]
\nonumber\\
&\times &
\left\{\exp\left[\frac{\sqq(r_1,s)\,\sqq(r_2,s)}{16\pi B(s)}\,T_A(b)\right] - 1\right\}\,.
\label{inc}
\EA 
%
The elastic cross section of a heavy quarkonium on a nucleon is rather small and the exponential in the last row of Eq.~(\ref{inc}) can be expanded. Then we arrive at a simple result \cite{Ivanov:2002kc},
%
\BA 
\frac{d^2\sigma_A^{inc}(s,b)}{d^2b}
\Biggr|_{l_c \gg R_A} 
&\approx&
\frac{T_A(b)}{16 \pi B(s)}\,
\Biggl |
\int d^2r\int_0^1 d\alpha\,
\Psi_{V}^{*}(\vec r,\alpha)\,
\Psi_{\gamma}(\vec r,\alpha)\,
\sqq(r,s)\,
\exp\left[-\frac{1}{2} \sqq(r,s)\, T_A(b)\right]\,
\Biggr |^2
\nonumber\\
&\equiv&
\frac{T_A(b)}{16 \pi B(s)}\,
\Biggl |
\int d^2r\int_0^1 d\alpha\,
\Sigma_{A}^{inc}(r,\alpha,s,b)\,
\Biggr |^2.
\label{inc-lcl}
 \EA
%

%
%
%
\subsection{Real part of the production amplitude and spin rotation effects}
\label{sub-sr}
%
%
%

The both Eqs.~(\ref{coh-lcl}) and (\ref{inc-lcl}) contain a small correction due to 
the real part of the $\gamma N\to V N$ amplitude applying 
the following replacement \cite{Bronzan:1974jh,Nemchik:1996cw,Forshaw:2003ki},
%
\BA
\sqq(r,s)
\Rightarrow
\sqq(r,s)
\,
\left(1 - i\,\frac{\pi}{2}\,\frac
{\partial
 \,\ln\,{\sqq(r,s)}}
{\partial\,\ln s} \right)\ .
  \label{re/im}
\EA
%
\\

The advantage of the $S$-wave heavy quarkonia, considered in the present paper, is based on a simple factorization of the radial and spin-dependent components of their wave functions. Here is well defined in the $Q\bar Q$ rest frame and can be obtained by solving the Schr\"odinger equation for various realistic interaction potentials between $Q$ and $\bar Q$ proposed in the literature. 
In our calculations, we choose two of them, the power-like (POW) \cite{Martin:1980jx,Barik:1980ai} and Buchm\"uller-Tye (BT) \cite{Buchmuller:1980su} potentials, which provide the best description of available data on charmonium electroproduction off protons, as was demonstrated in Ref.~\cite{Cepila:2019skb}.
For the dipole cross section, we employ three popular parametrizations - Kopeliovich-Sch\"afer-Tarasov (KST) from 
Ref.~\cite{Kopeliovich:1999am},  
Golec-Biernat-W\"usthoff (GBW) from
Refs.~\cite{GolecBiernat:1998js,GolecBiernat:1999qd} and 
Bartels-Golec-Biernat-Kowalski (BGBK) from Ref.~\cite{Bartels:2002cj}.

Treating the structure of the $V\to Q\bar Q$ vertex from Refs.~\cite{Ivanov:2002kc,Ivanov:2002eq,Ivanov:2007ms,Krelina:2018hmt,Cepila:2019skb,Kopeliovich:2022jwe}
the Melosh spin transformation \cite{Melosh:1974cu} is incorporated performing following substitutions in Eqs.~(\ref{coh-lcl}) and (\ref{inc-lcl}),
%
\BA
\Sigma_{A}^{coh}(r,\alpha,s,b)
\Rightarrow
\Sigma_{A}^{coh}(r,\alpha,s,b)\cdot 
     \Bigl [\Sigma^{(1)}(r,\alpha) + \Sigma^{(2)}(r,\alpha) \Bigr ]
\nonumber\\
\Sigma_{A}^{inc}(r,\alpha,s,b)
\Rightarrow
\Sigma_{A}^{inc}(r,\alpha,s,b)\cdot
     \Bigl [\Sigma^{(1)}(r,\alpha) + \Sigma^{(2)}(r,\alpha) \Bigr ],
\label{sr-coh-inc}
\EA
%
where
%
\BA
  \Sigma^{(1,2)}(r,\alpha)
   &=&  
   N\,
   K_{0,1}(m_Q r) \int\limits_{0}^{\infty} dp_T\,p_T\,
   J_{0,1}(p_T r) \Psi_V (\alpha,p_T) \,
    \mathcal{R}^{(1,2)}(p_T)
\label{sigma-sr1}
\EA
%
with
%
\BA
\mathcal{R}^{(1)}(p_T) 
=
\frac{2\,m_Q^2(m_L+m_T)+m_L\,p_T^2}{ m_T (m_L + m_T)} \,,
\qquad\qquad\quad
\mathcal{R}^{(2)}(p_T) 
&=&
\frac{m_Q^2(m_L+2m_T)-m_T\,m_L^2}{m_Q\,m_T (m_L+m_T)}\, p_T\,.
\label{sigma-sr2}
\EA
%
Here $N=Z_Q\,\sqrt{2 N_c \,\alpha_{em}}/4\,\pi$, where the factor $N_c=3$ represents the number of colors in QCD, $Z_Q=\frac{2}{3}$ and $\frac{1}{3}$  
are the charge-isospin factors for the production of charmonia and bottomonia, respectively, and
$J_{0,1}$ are the Bessel functions of the first kind.
The variables $m_{T,L}$ in the above formulas have the following form,
%
\BE
m_T = \sqrt{m_Q^2 + p_T^2} \,, 
\qquad m_L = 2\, m_Q\,\sqrt{\alpha(1-\alpha)}\,.
\EE
%

Note that the new form of Eqs.~(\ref{sr-coh-inc})-(\ref{sigma-sr2})
does not require performing the so-called {``\it resummation procedure''} proposed in \cite{Ivanov:2002kc} in order to include properly the spin rotation effects in 
nuclear photo-production cross sections (\ref{coh-lcl}) and (\ref{inc-lcl}).

Fir numerical calculations, following the results from Refs.~
\cite{Ivanov:2002kc,Ivanov:2002eq,Ivanov:2007ms,Krelina:2018hmt,Cepila:2019skb},
we relied on the charm and the bottom quark masses, corresponding to the values used with  the realistic phenomenological models for the $Q-\bar Q$ interaction potential, such as POW and BT.
Consequently, the LF quarkonium wave functions $\Psi_V(\alpha,r)$ have been obtained adopting the Lorentz boosting procedure as described in Ref.~\cite{Terentev:1976jk} and justified in Ref.~\cite{Kopeliovich:2015qna}.

%
%
%
\subsection{Corrections for a finite coherence length and the gluon shadowing}
\label{fcor}
%
%
%

As was already mentioned above, the Green function approach allows to include directly the effects of quantum coherence without any restrictions for the magnitude of CL, Eq.~(\ref{lc}). However, as an alternative and a more simple way,
instead of such a complicated method, one can use expressions ({\ref{coh-lcl}) and (\ref{inc-lcl})) for nuclear cross sections in the limit of long CL, $l_c\gg R_A$, 
and then provide additional corrections for a finite CL when $l_c\lsim R_A$. 
Such an incorporation of finite-$l_c$ effects via the effective correction factors (form factors), $F^{coh}(s,l_c)$ and $F^{inc}(s,l_c)$ 
based on VDM has been suggested in  Ref.~\cite{Hufner:1996jw} and employed in Ref.~\cite{Ivanov:2002kc} for calculations of charmonium photo-production off nuclei. 
Then nuclear cross sections corrected to the finite coherence length effects are given as 
%
\BA
\frac{d^2
  \sigma_A^{coh}(s,b)}{d^2 b} 
  =
\frac{d^2
  \sigma_A^{coh}(s,b)}{d^2 b}\Biggl|_{l_c\gg R_A} \cdot F^{coh}\bigl (s,l_c(s)\bigr )\,,
  \qquad
\frac{d^2   
  \sigma_A^{inc}(s,b)}{d^2 b} 
=
\frac{d^2
  \sigma_A^{inc}(s,b)}{d^2 b}\Biggr|_{l_c\gg R_A} \cdot F^{inc}\bigl (s,l_c(s)\bigr )\,.
\EA
%

In the present paper, as the further improvement, instead of VDM 
the corresponding factors $F(s,l_c)$
have been calculated within a more sophisticated Green function formalism and
have the following form,
%
\BA
  F^{coh}(s,l_c) &=& \int\!d^2b
    \left|
    \,\int\limits_{-\infty}^\infty\!dz\,\rho_A(b,z)\,F_1(s,b,z,l_c)\,
    \right|^2 \bigl / \,\Bigl (...\Bigr )\Bigl |_{l_c \to \infty} \,, 
    \label{fcoh}
    \\
  F^{inc}(s,l_c) &=& \int\!d^2b
    \,\int\limits_{-\infty}^\infty\!dz\,\rho_A(b,z)
    \,\Bigl |F_1(s,b,z,l_c)-F_2(s,b,z,l_c)\Bigr |^2
    \bigl / \,\Bigl (...\Bigr )\Bigl |_{l_c \to \infty} \,,
    \label{finc}
\EA
%
where the functions $F_1$ and $F_2$ read,
%
 \BA
F_1(s,b,z,l_c) &=&
\int\limits_0^1 d\alpha
\int d^{2} r_{1}\,d^{2} r_{2}\,
\Psi^{*}_{V}(\vec r_{2},\alpha)\,
G_{Q\bar Q}(z^\prime,\vec r_{2};z,\vec r_{1};l_c)\,
\sqq(r_{1},s)\,
\Psi_{\gamma}(\vec r_{1},
\alpha)
\Bigl|_{z^\prime\to\infty}
\label{f1}
\\
F_{2}(s,b,z,l_c) &=& \frac{1}{2}\,
\int\limits_{-\infty}^{z} dz_{1}\,\rho_{A}(b,z_1)\,
\int\limits_0^1 d\alpha\int d^2 r_1\,
d^2 r_{2}\,d^2 r\,
\Psi^*_V (\vec r_2,\alpha)\nonumber \\
&\times&
G_{Q\bar Q}(z^{\prime}\to\infty,\vec r_2;z,\vec r;l_c)\,
\sqq(\vec r,s)\,
G_{Q\bar Q}(z,\vec r;z_1,\vec r_1;l_c)\,
\sqq(\vec r_1,s)\,
\Psi_{\gamma}(\vec r_1,\alpha)\, .
\label{f2}
 \EA
%

Here $\Psi_{\gamma}(\vec r,\alpha) = K_0(m_Q r)$, and the Green function 
$G_{Q\bar Q}(z^\prime,\vec r_2;z,\vec r_1;l_c)$ describes the propagation of an interacting $Q\bar Q$ pair in a nuclear medium between points with longitudinal coordinates $z$ and $z^\prime$ and with initial and final separations $\vec r_1$ and $\vec r_2$.
In calculations for the sake of simplicity, we employed the quadratic form for the dipole cross section $\sqq(r,s) = C(s)\,r^2$,
and the harmonic oscillatory form for the LF $Q-\bar Q$ interaction potential in the evolution equation for the Green function (e.g. see Ref.~\cite{Kopeliovich:2001xj}). 
Another simplification is related to
a constant nuclear density, $\rho_A(b,z) = \rho_0 \Theta (R_A^2 - b^2 - z^2)$, which is rather accurate for heavy nuclei used in our analysis. Consequently, for the LF quarkonium wave functions, we obtained the following Gaussian shape \cite{Kopeliovich:2001xj} for the $1S$ and $2S$ states,
%
\BA
\Psi_V(r,\alpha) 
&=& 
C_V\,a^2(\alpha)\,f(\alpha)\,\exp\Bigl [- \frac{1}{2} a^2(\alpha) r^2\Bigr ]
\\
\Psi_{V^{\prime}}(r,\alpha) 
&=& 
C_{V^{\prime}}\,a^2(\alpha)\,f(\alpha)\,\exp\Bigl [- \frac{1}{2} a^2(\alpha) r^2\Bigr ]
\Bigl \{1 + 4 h(\alpha) - \beta\, 2 a^2(\alpha) r^2 \Bigr \}\,,
\label{psi-lc}
\EA
%
where
%
\BE
f(\alpha) = \exp\Bigl [ - h(\alpha)\Bigr ]
          =
\exp\biggl [ - \frac{m_Q^2}{2 a^2(\alpha)}
                        + \frac{4 \alpha (1-\alpha) m_Q^2}{2 a^2(\alpha)} \biggr ]\, ,
\label{fa}
\EE
%
and the parameter $\beta$, controlling the position of the node, has been determined from the orthogonality condition $\int d^2 r\,d \alpha \Psi_V(r,\alpha)\,\Psi_{V^{\prime}}(r,\alpha)= \delta_{VV^{\prime}}$. We have found $\beta = 0.908$ and $0.963$ for production of $\psip$ and $\Yp$, respectively. The function $a^2(\alpha) = 2\alpha(1-\alpha)\,m_Q\,\omega$, where the oscillatory frequency $\omega = (\left.M_{V^\prime}\right. - \left.M_V\right.)/2\approx 0.3\,\GeV$.

The above approximations substantially simplify the calculations of the $l_c$-correction factors $F^{coh}$ and $F^{inc}$
since allow to obtain an explicit analytical harmonic oscillatory form for the Green function \cite{fg},  
%
 \BA
G_{Q\bar Q}(\vec{r_2},z_2;\vec{r_1},z_1;l_c) 
=
\frac{b(\alpha)}{2 \pi i\,
{\rm sin}(\Omega\Delta z)}\, {\rm exp}
\left\{
\frac{i\,b(\alpha)}{{2\,\rm sin}(\Omega \Delta z)}\,
\Bigl[(r_1^2+r_2^2)\,{\rm cos}(\Omega \Delta z) -
2 \vec{r_1}\cdot\vec{r_2}\Bigr]\right\}\,
{\rm exp}\left[-
\frac{i \Delta z}
{l_c}\right]\,,
\label{GF-HO}
 \EA
with $\Delta z = z_2-z_1$, $l_c = 2 k \alpha (1-\alpha)/m_Q^2$ and
%
\BA
\Omega
=      
\frac{b(\alpha)}{k\, \alpha (1 - \alpha)}
&=&
\frac{
\sqrt{a^4(\alpha) - i\,k\,\alpha\,(1-\alpha)\, 
C_{eff}(s,\alpha)\,\rho_A({b},z_2)
}}
{k\,\alpha\,(1 - \alpha)}\,.
\label{GF-HO-s}
\EA
%
Here,
considering the standard saturated shape of the dipole cross section,
%
\BA
\sqq(\vec r,s) =
\sigma_0\,
\biggl (1 - \exp \Bigl [ - \frac{r^2}{r_0^2(s)}\Bigr ] \biggr )
\,,
\label{gbw}
\EA
the factor $C_{eff}$ in Eq.~(\ref{GF-HO-s}) can be expressed in the following form,
%
\BA
C_{eff}(s,\alpha) 
= 
C(s)\, \biggl\{1 - \exp\biggl [-\,\frac{1} { a^2(\alpha)\,r_0^2(s)}\,\biggr ] \biggr \}\, a^2(\alpha)\,r_0^2(s)\,,
\qquad\qquad\qquad
C(s) = \sigma_0/r_0^2(s)\,.
\EA

\BF
\hspace*{0.00cm}
\PSfig{1.30}{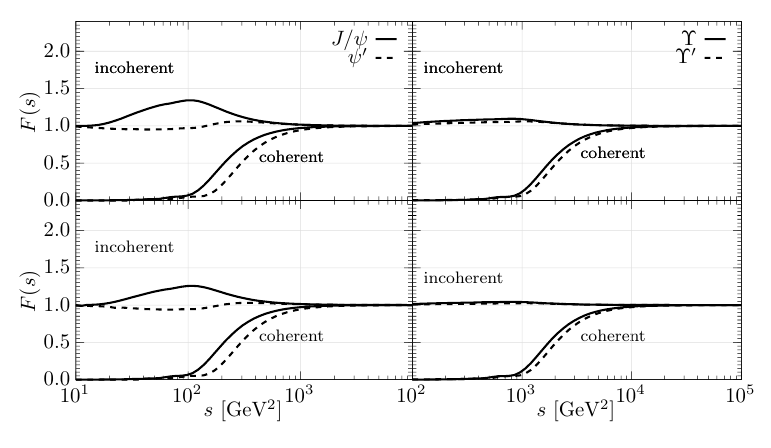}
\vspace*{-0.6cm}
\Caption
   {
  \label{Fig-lc-gf}
  (left panels) - 
  the $l_c$-correction factors for coherent and incoherent production of $\Jpsi$ (solid lines) and $\psip$ (dashed lines) in $Pb-Pb$ UPC obtained within the color dipole approach based on the Green function technique, 
  Eqs.~(\ref{fcoh})-(\ref{f2}). The quarkonium wave functions are generated by the BT potential.
  (right panels) -
  the same as left panels but for production of $\Y$ and $\Yp$.
  The top and bottom panels correspond to calculations using KST and GBW models for the dipole cross section, respectively.
  }
\EF

The $l_c$-correction factors $F(s)$, obtained from Eqs.~(\ref{fcoh})-(\ref{f2}),
are depicted in Fig.~\ref{Fig-lc-gf} 
as a function of the square of c.m. energy $s$
for coherent and incoherent $\Jpsi$ (solid lines) and $\psip$ (dashed lines) photo-production on the lead target (left panels).
Analogous results for $F(s)$ for photo-production of $\Y$ and $\Yp$ are depicted on the right panels of the same Fig.~\ref{Fig-lc-gf}. 
Here the top and bottom panels correspond to calculations using KST and GBW parametrization for the dipole cross section, respectively.

One can see from Fig.~\ref{Fig-lc-gf} that the effects of a finite CL are important for 
energies $s\lsim 10^3\,\GeV^2$ and $s\lsim 10^4\,\GeV^2$ in the production of charmonia and bottomonia, respectively. This is a direct consequence of the CL dependence on the quarkonium mass as given by Eq.~(\ref{lc}).
Figure~\ref{Fig-lc-gf} also demonstrates that contraction of the CL at smaller values of $s$ leads to a significant reduction of the coherent cross sections for the $1S$ quarkonium states. However, the corresponding incoherent cross sections are enhanced  by $\sim 20\%-30\,\%$ and $\sim 7\%-12\,\%$ for production of $\Jpsi(1S)$ and $\Y(1S)$, respectively.  

For radially excited $2S$ quarkonia, the manifestations of finite-$l_c$ 
is affected by the nodal structure of quarkonium wave functions. Its influence is stronger for  $\psip(2S)$ in comparison with $\Yp(2S)$, leading to a more complicated non-monotonic behaviour of the factor $F^{inc}(s)$ at small values of $s$. On the other hand, a stronger energy dependence from the region of small $Q\bar Q$ transverse separations below the node position compared to large $Q\bar Q$ dipole sizes above the node position causes a weakening of the node effect with rising energy, resulting in a gradual convergence of factors $F^{inc}_{V^{\prime}}(s)$ and $F^{inc}_{V}(s)$ 
towards large $s$.

\vspace*{0.2cm}
The leading twist gluon shadowing was introduced within the dipole representation in \cite{Kopeliovich:1999am} and applied to photo-production of vector mesons on nuclei in  Refs.~\cite{Kopeliovich:2001xj,Ivanov:2002kc,Nemchik:2002ug,Kopeliovich:2007wx}.
In the present paper, we include only  one gluon Fock state $|Q\bar QG\ra$,  because
higher multi-gluon components give negligible contribution to nuclear shadowing within the 
kinematic regions of present UPC experiments at the LHC (see analysis and discussion in Ref.~\cite{Kopeliovich:2022jwe}).
Since the dipole cross section $\sqq(r,s)$ at small dipole sizes $\vec r$
depends on the gluon distribution in the target, nuclear shadowing of the
gluon distribution can be interpreted within the parton model as a reduction of $\sqq(r,s)$ in nuclear reactions with respect to processes on the nucleon,
%
\BE
  \sqq(r,x) \Rightarrow \sqq(r,x) \cdot R_G(x,b)\,.
\label{eq:dipole:gs:replace}
\EE
%
Here the Gribov correction factor $R_G(x,b)$, related to the $Q\bar QG$ component of the photon, was calculated at impact parameter $b$ using the Green function
formalism \cite{Kopeliovich:1999am,Kopeliovich:2001xj,Kopeliovich:2001ee,Ivanov:2002kc,Nemchik:2002ug,Kopeliovich:2008ek,Krelina:2020ipn}\footnote{An alternative estimation of gluon shadowing effects in charmonium production in UPC can be found in \cite{guzey-1,guzey-2}, for example.}
(see also Fig.1 and a discussion in Ref.~\cite{Kopeliovich:2022jwe}).

\BF
\PSfig{1.1}{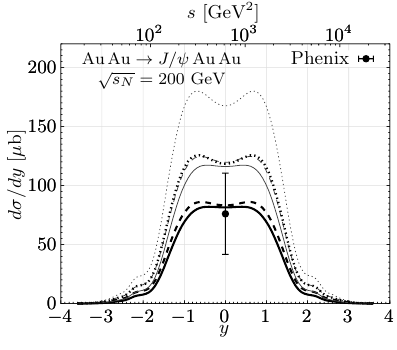}~~~~
\PSfig{1.1}{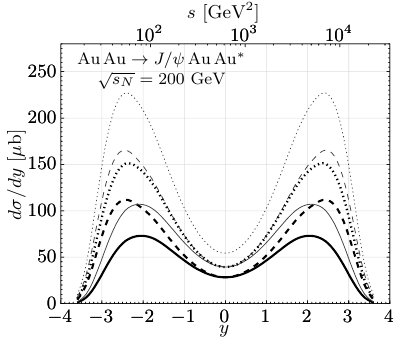}~~~~\\
\PSfig{1.1}{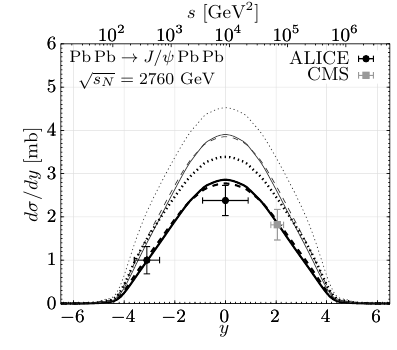}~~~~
\PSfig{1.1}{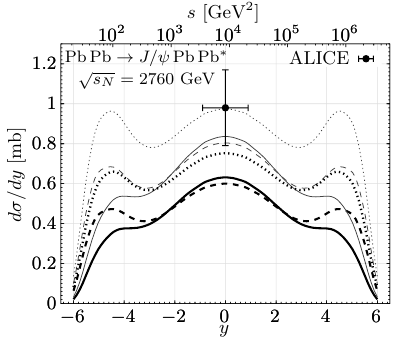}~~~~\\
\PSfig{1.1}{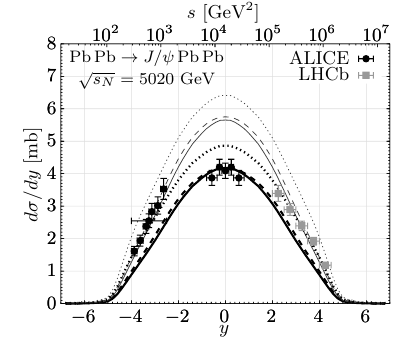}~~~~
\PSfig{1.1}{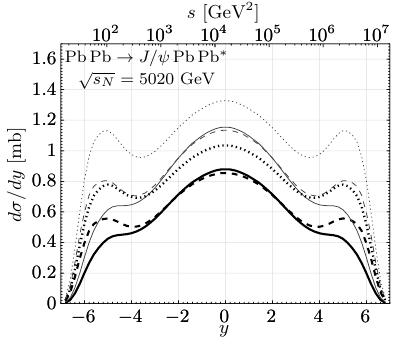}~~~~\\
\vspace*{-0.30cm}
\Caption{
\label{Fig-UPC1psi}
  Rapidity distributions of coherent (left panels) and incoherent (right panels) charmonium
  photo-production in UPC at RHIC collision energy $\sqrt{s_N}=200\,\GeV$ (top panels) and at LHC energies $\sqrt{s_N} = 2.76\,\TeV$ (middle panels) and $\sqrt{s_N} = 5.02\,\TeV$ (bottom panels). The nuclear cross sections are calculated with charmonium wave functions generated by the POW (thin lines) and BT (thick lines) potentials and with  GBW (solid lines), KST (dashed lines) and BGBK (dotted lines) models for the dipole cross section. The data are taken from
  PHENIX \cite{Afanasiev:2009hy},  
  CMS \cite{Khachatryan:2016qhq}, ALICE \cite{Abelev:2012ba,Abbas:2013oua,Adam:2015sia,Acharya:2019vlb,ALICE:2021gpt} and  LHCb \cite{LHCb:2018ofh,LHCb:2022ahs} collaborations.
  }
\EF

%
%
%
\section{Model predictions vs available data}
\label{res}
%
%
%

We have calculated rapidity distributions $d\sigma/dy$ for
the coherent and incoherent heavy quarkonium photo-production in UPC according to Eq.~(\ref{cs-upc}). 
Here the $|Q\bar Q\ra$ Fock component of the photon was treated in the ``frozen'' approximation,
$l_c\gg R_A$ (see Eqs.(\ref{coh-lcl}) and (\ref{inc-lcl})).
Moreover,
in order to calculate the nuclear cross sections (\ref{inc-lcl}) for the incoherent (quasi-elastic) production, one should know also the slope parameter for the elastic process $\gamma N\to V N$. Here we rely on the standard Regge form,
%
$B_{\Jpsi}(s)=B_0 + 2\,\alpha'(0) \ln\big(s/s_0\big)$,
%
where the parameters $\alpha'=0.171\GeV^{-2}$, the slope of the Pomeron trajectory, and
$B_{0}=1.54\,\GeV^{-2}$ were fitted in \cite{Cepila:2019skb} to data on $\Jpsi$ photo-production with $s_0=1\GeV^2$. 
The slope for $1S$ bottomonium photo-production was fitted to data in \cite{Cepila:2019skb} and found to have a smaller value than for $\Jpsi$,
$B_{\Y}(s)\approx B_{\Jpsi}(s) -1\GeV^{-2}$.
For the production of radially excited $2S$ state of bottomonia, the node effect is negligibly small, and one can safely use the same magnitudes of the slope parameter for both $1S$ and $2S$ states, i.e. $B_{\Yp}(s)\sim B_{\Y}(s)$. Not so for the production of radially excited charmonia where the difference of diffraction slopes $\Delta_B(s)=B_{\Jpsi}(s)-B_{\psip}(s)$ cannot be neglected. 
Here we adopt a parametrization of the factor $\Delta_B(s)$ 
from Ref.~\cite{Cepila:2019skb} (see also Ref.~\cite{Nemchik:1997xb}).

Besides nuclear suppression of the lowest Fock component, $|Q\bar Q\ra$,
we included in our predictions two main phenomena affecting the nuclear cross sections:
the gluon shadowing and the finite-$l_c$ corrections. Whereas the former dominates at large photon energies, the latter is prominent at smaller energies, 
when $l_c\lsim R_A$.
Here the $l_c$-correction factors are
calculated for the first time within a rigorous Green function formalism as described in Sect.~\ref{fcor}.

In Fig.~\ref{Fig-UPC1psi} we present 
our results for the rapidity distributions $d\sigma/dy$ of coherent (left panels) and incoherent (right panels) charmonium photo-production in UPC obtained for $\sqrt{s_N} = 200\,\GeV$ (top panels),
$\sqrt{s_N} = 2.76\,\TeV$ (middle panels) and $\sqrt{s_N} = 5.02\,\TeV$ (bottom panels).
Calculations have been performed for charmonium wave functions generated by two distinct $Q\bar Q$ potentials, POW (thin lines) and BT (thick lines).  
For the dipole cross sections $\sqq$ we adopted three different parametrizations,
GBW (solid lines), KST (dashed lines) and BGBK (dotted lines).
Here the model predictions are tested
by the RHIC data from the PHENIX experiment \cite{Afanasiev:2009hy},
by the LHC data from the CMS \cite{Khachatryan:2016qhq} and ALICE \cite{Abelev:2012ba,Abbas:2013oua,Adam:2015sia} collaborations at c.m. collision energy $\sqrt{s_N} = 2.76\,\TeV$, as well as by the 
ALICE \cite{Acharya:2019vlb,ALICE:2021gpt} and 
LHCb \cite{LHCb:2018ofh,LHCb:2022ahs} data at $\sqrt{s_N} = 5.02\,\TeV$.

In Fig.~\ref{Fig-UPC1psi} one can see that the values of $d\sigma/dy$ 
strongly correlate with the shape of the quarkonium wave functions determined with various models for $Q-\bar Q$ interaction potentials (compare thin and thick lines).
While in charmonium production in UPC, the POW (thin lines) and BT (thick lines) models lead to rather different predictions for $d\sigma/dy$, in the bottomonium case both models 
give quite similar (almost identical) results (see Fig.~\ref{Fig-UPC1y}). This is in correspondence with our previous studies \cite{Cepila:2019skb} of quarkonium electroproduction off protons.

\BF
\PSfig{1.1}{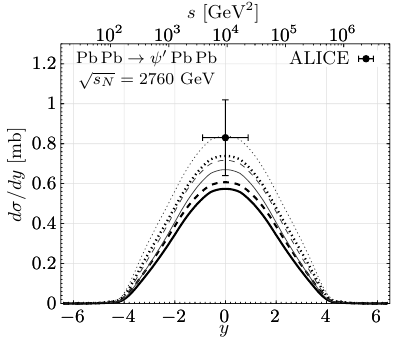}~~~~
\PSfig{1.1}{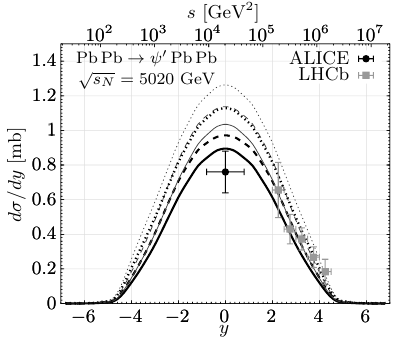}~~~~\\
\vspace*{-0.30cm}
\Caption{
  \label{Fig-UPC1psi2S}
  The same as Fig.~\ref{Fig-UPC1psi} but for the coherent $\psip(2S)$ production in UPC
  at the collision energy $\sqrt{s_N} = 2.76\,\TeV$ (left panel) and $\sqrt{s_N} = 5.02\,\TeV$ (right panel).
  The experimental values are taken from the ALICE
  \cite{Adam:2015sia,ALICE:2021gpt} and LHCb \cite{LHCb:2022ahs} collaborations.
  }
\EF

The experimental data on the production of radially excited heavy quarkonia in UPC are very scarcy. The ALICE collaboration \cite{Adam:2015sia} measured $d\sigma/dy$ for coherent production of $\psip(2S)$ at $\sqrt{s_N} = 2.76\,\TeV$ and $y=0$, as is depicted in the left panel of Fig.~\ref{Fig-UPC1psi2S} together with our results. Besides, a new experimental value of $d\sigma/dy$ has been obtained recently at higher energy $\sqrt{s_N} = 5.02\,\TeV$  in the ALICE experiment \cite{ALICE:2021gpt} at $y=0$, as well as several data points of $d\sigma/dy$ by the LHCb \cite{LHCb:2022ahs} experiment at various positive rapidities (see the right panel of Fig.~\ref{Fig-UPC1psi2S}).
One can see a reasonable agreement of our calculations with these data for both $c-\bar c$ interaction potentials, as well as for all models of the dipole cross section. Whereas at $\sqrt{s_N} = 2.76\,\TeV$ the better agreement with data is achieved with charmonium wave functions generated by the POW $c-\bar c$ potential, the higher collision energy  $\sqrt{s_N} = 5.02\,\TeV$ prefers excited charmonia described by the BT potential model. 

\BF
\PSfig{1.1}{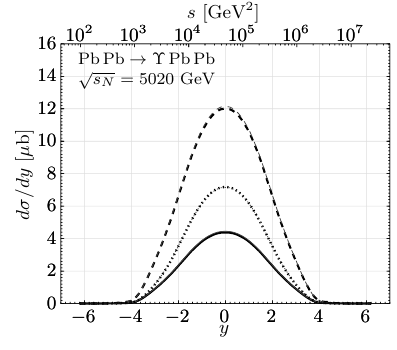}~~~~
\PSfig{1.1}{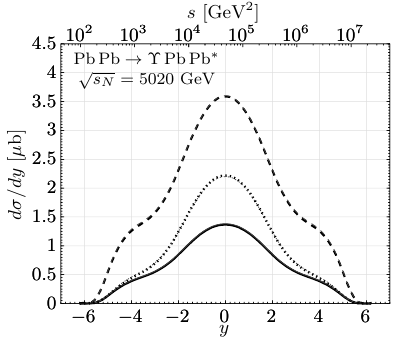}~~~~\\
\vspace{-0.30cm}
\Caption{
  \label{Fig-UPC1y}
  The same as Fig.~\ref{Fig-UPC1psi} but for the bottomonium production in UPC at the collision energy $\sqrt{s_N} = 5.02\,\TeV$.
  }
\EF

The next Fig.~\ref{Fig-UPC1y} represents analogous predictions as Fig.~\ref{Fig-UPC1psi} but for production of bottomonia in UPC. Here the GBW, KST and BGBK parametrizations give rather different results in correspondence with the previous analysis of the process $\gamma N\to\Y N$ in Ref.~\cite{Cepila:2019skb}. 
Differences in predictions using various models for $\sqq(r)$ can be treated as a measure of the theoretical uncertainty in our results.

\BF
\PSfig{1.1}{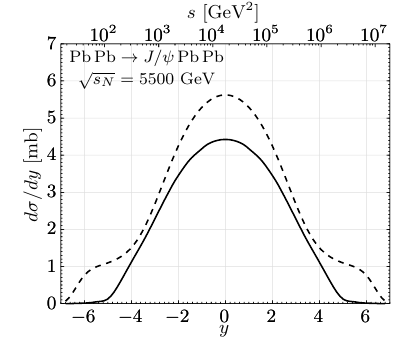}~~~~
\PSfig{1.1}{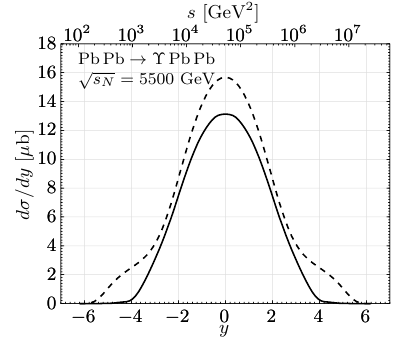}~~~~\\
\PSfig{1.1}{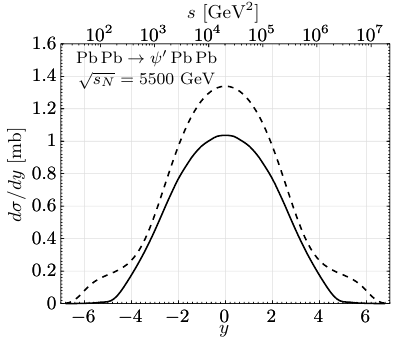}~~~~
\PSfig{1.1}{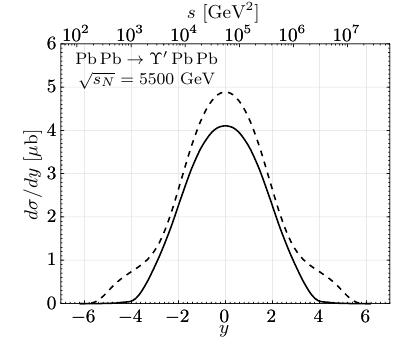}~~~~\\
\vspace*{-0.3cm}
\Caption{
  \label{Fig-UPC2psi-y}
  Manifestations of particular nuclear effects in
  coherent charmonium (left panels) and bottomonium (right panels)
  photo-production in UPC at the LHC collision energy $\sqrt{s_N}= 5.5\,\TeV$. Here top and bottom panels correspond to the production of 1S and 2S quarkonium states, respectively. The nuclear cross sections are calculated with charmonium wave functions generated by the BT potential adopting the KST model for the dipole cross section. 
  The dashed lines represent predictions in the high energy eikonal limit, Eq.~(\ref{coh-lcl}).
  The solid lines include additional corrections for a finite CL and the gluon shadowing.
  }
\EF

Figure~\ref{Fig-UPC2psi-y} demonstrates importance of particular nuclear effects for the rapidity dependence  $d\sigma/dy$
at energy $\sqrt{s_N}= 5.5\,\TeV$ in coherent production of charmonia (left panels) and bottomonia (right panels) in UPC, whereas the top and bottom panels represent model predictions for 1S and 2S quarkonium states, respectively.
Here dashed lines correspond to our calculations in the standard high energy eikonal limit, $l_c\gg R_A$,
Eq.(\ref{coh-lcl}).
The solid lines incorporate additionally two effects,  
corrections to a finite CL and the leading twist GS effect. 
Our model predictions are based on the KST model for the dipole cross section and include quarkonium wave functions determined from the BT model for the $Q-\bar Q$ interaction potential.

One can see from Fig.~\ref{Fig-UPC2psi-y} that differences between solid and dashed curves at large forward (backward) rapidities represent the relevance of the finite-$l_c$ corrections which is more pronounced in coherent production of bottomonia compared to the charmonium case. Here effects of gluon shadowing are substantially diminished due to their small contribution to $d\sigma/dy$ at large photon energies given by a significantly reduced photon flux, Eq.~(\ref{flux}).
However, the latter effects dominate at the mid rapidities, where effects of CL reduction do not play any role since the ``frozen'' eikonal limit is acquired for the higher twist shadowing correction.  Both effects substantially suppress the cross sections $d\sigma/dy$. In comparison to charmonium production, a weaker onset of gluon shadowing at a fixed $\sqrt{s_N}$ in production of bottomonia is caused by larger values of the both the variable $x$ and the corresponding scale $\propto M_{\Y}^2\gg M_{\Jpsi}^2$ (see also Fig.1 in Ref.~\cite{Kopeliovich:2022jwe}).

\BF
\PSfig{1.1}{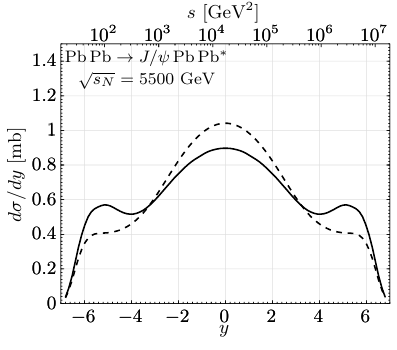}~~~~
\PSfig{1.1}{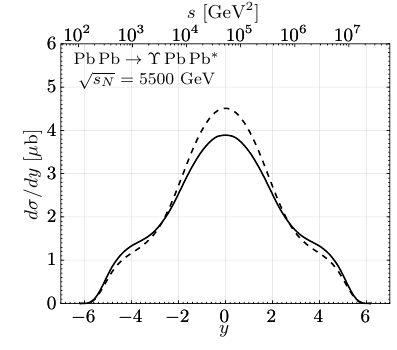}~~~~\\
\PSfig{1.1}{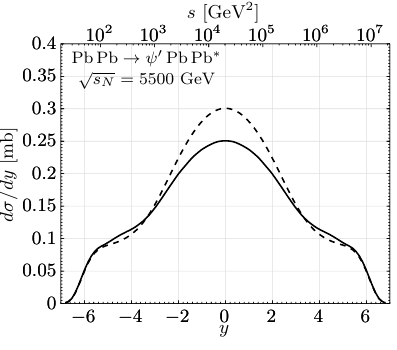}~~~~
\PSfig{1.1}{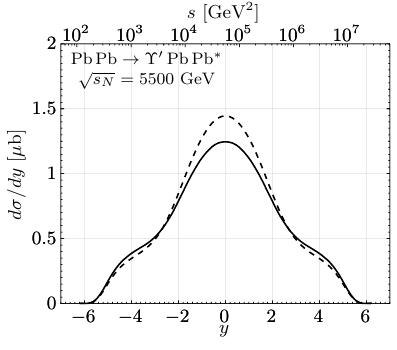}~~~~\\
\vspace*{-0.3cm}
\Caption{
  \label{Fig-UPC3-inc}
  The same as Fig.~\ref{Fig-UPC2psi-y} but for incoherent production of the ground state $1S$ (top panels) and radially excited $2S$ (bottom panels) quarkonia in UPC at $\sqrt{s_N}=5.5\,\TeV$.
  }
\EF

The last Fig.~\ref{Fig-UPC3-inc} illustrates a manifestation of particular nuclear effects 
at energy $\sqrt{s_N}=5.5\,\TeV$
in the incoherent production of the $1S$ ground state (top panels) and $2S$ radially excited (bottom panels) quarkonia. 
The effects of reduced CL are visible only in the production of $\Jpsi$ at large positive and negative rapidities. For other quarkonium states, they can be neglected in accordance with values of $l_c$-correction factors which are close to unity (see Fig.\ref{Fig-lc-gf}).
At the mid rapidity, the
difference between solid and dashed lines shows the net effect of gluon shadowing with a maximal magnitude at $y=0$.

%
%
%
\section{Conclusions}
\label{conclusions}
%
%
%

In this paper we treat the heavy quarkonium [$\Jpsi(1S),\psip(2S),\Y(1S),\Yp(2S)$]
production in heavy-ion UPC 
within the light-front QCD dipole approach
in the energy range accessible by experiments at RHIC and the LHC. Here the main observations are the following:

\begin{itemize}

\item 
The quarkonium wave functions are well defined in the $Q\bar Q$ rest frame. They have been included in our calculations by solving the Schr\"odinger equation with several realistic $Q-\bar Q$ interaction potentials.
Consequently, the corresponding LF wave functions have been generated performing the boosting to the LF frame using the so-called Terent'ev prescription from Ref.~\cite{Terentev:1976jk}, which was found in Ref.~\cite{Kopeliovich:2015qna} to have a reasonable accuracy in comparison with exact solution. 

Here we ignore the model for the photon-like $V\to Q\bar Q$ transition, frequently used in the literature,  
avoiding so too large weight of the $D$-wave component in the $Q\bar Q$ rest frame, in consistence with solutions of the Schr\"odinger
equation.

\item 
The spin-dependent part of the wave function for $S$-wave quarkonia can be safely factorized from the radial component. 
Consequently, we
perform explicitly the transformation of two-dimensional heavy (anti)quark spinors from the rest to the LF frame known as the Melosh spin rotation. We derived new formulas for coherent and incoherent nuclear cross sections incorporating such a transformation
(see Eqs.~(\ref{sr-coh-inc})-(\ref{sigma-sr2})).

\item
UPC  at RHIC and the LHC at the mid rapidities, provide a sufficiently high energy and a long  coherence length for the lowest $|Q\bar Q\ra$ Fock state of the photon. As far as CL  considerably exceeds the nuclear 
size, one can rely on the high-energy eikonal approximation for nuclear effects (see Eqs.~(\ref{coh-lcl}) and (\ref{inc-lcl})).
The corresponding shadowing correction is small since diminishes with heavy quark mass as $1/m_Q^2$, so represents the higher twist effect. 

\item
At forward and/or backward rapidities, the eikonalization of $\sqq(r,s)$ cannot be applied anymore and 
we included corrections for a finite coherence length which have been calculated for the first time within a rigorous quantum-mechanical description, summing up all possible paths of the quarks.
A proper treatment of finite-$l_c$ corrections is absent in the most of the recent calculations of UPC.
These corrections lead to a significant modification
of rapidity distributions $d\sigma/dy$ at small photon energies when $l_c\lsim R_A$ (see Figs.~\ref{Fig-lc-gf}, \ref{Fig-UPC2psi-y} and \ref{Fig-UPC3-inc}).

\item
We also included the gluon shadowing corrections related to higher Fock components of the photon containing gluons. Those components have a coherence length much shorter than the lowest 
$|Q\bar Q\ra$ Fock state.
They represents the leading-twist effect since the transverse size of the $Q\bar Q-G$ dipole is much larger compared to the small-sized $Q\bar Q$ fluctuation and is almost independent of $m_Q$.
The dominant contribution to nuclear shadowing comes from the $|Q\bar QG\ra$ Fock state of the photon. 
In calculations of the corresponding shadowing factor one cannot use the standard eikonal approximation even at high energies since the $Q\bar Q-G$ size fluctuates during propagation through the medium. This is why we applied the Green function formalism.
The higher photon fluctuations with more gluons do not cause a significant shadowing effect (see also Sec. IV in Ref.~\cite{Kopeliovich:2022jwe}).

\item
We have also studied differences in our predictions employing KST, GBW and BGBK phenomenological parametrizations for the dipole cross section $\sqq(r,s)$ in order to estimate 
a corresponding measure of the theoretical uncertainty in our current analysis.
We  concluded that whereas in charmonium production in UPC, the main source of theoretical uncertainties is related predominantly to our choice of quarkonium wave functions, in the bottomonium case, the main role in the variability of predictions is played by various models for $\sqq(r,s)$.

\item
Our predictions for $d\sigma/dy$ are in a rather good accord with available data on coherent production of $\Jpsi(1S)$ and $\psip(2S)$ in UPC at the energies of RHIC and LHC (see Figs.~\ref{Fig-UPC1psi} and \ref{Fig-UPC1psi2S}). 
They can be tested not only by measurements at the LHC, but also in future experiments at the planned electron-ion colliders.

\end{itemize}

\begin{acknowledgements}
This work was supported in part by grant ANID PIA/APOYO AFB220004.
The work of J.N. was partially supported by Grant
No. LTT18002 of the Ministry of Education, Youth and
Sports of the Czech Republic,
by the project of the
European Regional Development Fund No. CZ.02.1.01/0.0/0.0/16\_019/0000778
and by the Slovak Funding Agency, Grant No. 2/0020/22.
The work of M.K. was supported by the project of the International Mobility of Researchers - MSCA IF IV at CTU in Prague 
CZ.02.2.69/0.0/0.0/20\_079/0017983, Czech Republic.
\end{acknowledgements}



\begin{thebibliography}{99}

\bibitem{Hufner:1997jg}
   J.~Hufner and B.~Z.~Kopeliovich,
   ``J / Psi N and Psi-prime N total cross-sections from photo-production data: Failure of vector dominance,''
   Phys. Lett. B \textbf{426}, 154-160 (1998).

\bibitem{Kopeliovich:1991pu} 
   B.Z.~Kopeliovich and B.G.~Zakharov;
   ``Quantum effects and color transparency in charmonium photoproduction on nuclei,''
   Phys. Rev. D \textbf{44}, 3466 (1991).

\bibitem{Ivanov:2007ms}
   Y.~Ivanov, B.~Kopeliovich and I.~Schmidt,
   ``Vector meson production in ultra-peripheral collisions at LHC,''
   arXiv:{0706.1532 [hep-ph]}.

\bibitem{Kopeliovich:1993pw} 
   B.Z.~Kopeliovich, J.~Nemchik, N.N.~Nikolaev and B.G.~Zakharov;
   ``Decisive test of color transparency in exclusive electroproduction of vector mesons,''
   Phys. Lett. B \textbf{324}, 469 (1994).

\bibitem{Nemchik:1996pp}
   J.~Nemchik, N.~N.~Nikolaev, E.~Predazzi and B.~Zakharov,
   ``Color dipole systematics of diffractive photoproduction and electroproduction of vector mesons,''
   Phys. Lett. B \textbf{374}, 199 (1996).

\bibitem{Nemchik:1996cw}
   J.~Nemchik, N.~N.~Nikolaev, E.~Predazzi and B.~Zakharov,
   ``Color dipole phenomenology of diffractive electroproduction of light vector mesons at HERA,''
   Z. Phys. C \textbf{75}, 71 (1997).

\bibitem{Kopeliovich:1993gk}
   B.~Kopeliovich, J.~Nemchik, N.~N.~Nikolaev and B.~Zakharov,
   ``Novel color transparency effect: Scanning the wave function of vector mesons,''
   Phys. Lett. B \textbf{309}, 179 (1993).


\bibitem{Nemchik:1994fp}
   J.~Nemchik, N.~N.~Nikolaev and B.~Zakharov,
   ``Scanning the BFKL pomeron in elastic production of vector mesons at HERA,''
   Phys. Lett. B \textbf{341}, 228 (1994).

\bibitem{Kopeliovich:2001xj}
   B.~Kopeliovich, J.~Nemchik, A.~Schafer and A.~Tarasov,
   ``Color transparency versus quantum coherence in electroproduction of vector mesons off nuclei,''
   Phys. Rev. C \textbf{65}, 035201 (2002).

\bibitem{Kopeliovich:2007wx}
   B.~Kopeliovich, J.~Nemchik and I.~Schmidt,
   ``Production of Polarized Vector Mesons off Nuclei,''
   Phys. Rev. C \textbf{76}, 025210 (2007).


\bibitem{Nemchik:1994fq}
   J.~Nemchik, N.~N.~Nikolaev and B.~Zakharov,
   ``Anomalous a-dependence of diffractive leptoproduction of radial excitation rho-prime (2s),''
   Phys. Lett. B \textbf{339}, 194 (1994).

\bibitem{Hufner:2000jb} 
   J.~Hufner, Y.P.~Ivanov, B.Z.~Kopeliovich and A.V.~Tarasov;
   ``Photoproduction of charmonia and total charmonium proton cross-sections,''
   Phys. Rev. D \textbf{62}, 094022 (2000).

\bibitem{Nemchik:2000de}
   J.~Nemchik,
   ``Wave function of 2S radially excited vector mesons from data for diffraction slope,''
   Phys. Rev. D \textbf{63}, 074007 (2001).

\bibitem{Nemchik:2000dd}
   J.~Nemchik,
   ``Anomalous t dependence in diffractive electroproduction of 2S radially excited light vector mesons at HERA,''
   Eur. Phys. J. C \textbf{18}, 711 (2001).

\bibitem{Kopeliovich:2000ra} 
   B.Z.~Kopeliovich, J.~Raufeisen and A.V.~Tarasov,
   ``Nuclear shadowing and coherence length for longitudinal and transverse photons,''
   Phys. Rev. C \textbf{62}, 035204 (2000).

\bibitem{Kopeliovich:2022jwe}
   B.~Z.~Kopeliovich, M.~Krelina, J.~Nemchik and I.~K.~Potashnikova,
   ``Coherent photoproduction of heavy quarkonia on nuclei,''
   Phys. Rev. D \textbf{105}, 054023 (2022).

\bibitem{Ivanov:2002kc}
   Y.~Ivanov, B.~Kopeliovich, A.~Tarasov and J.~Hufner,
   ``Electroproduction of charmonia off nuclei,''
   Phys. Rev. C \textbf{66}, 024903 (2002).

\bibitem{Kopeliovich:1999am} 
   B.Z.~Kopeliovich, A.~Schafer and A.V.~Tarasov,
   ``Nonperturbative effects in gluon radiation and photoproduction of quark pairs,''
   Phys. Rev. D \textbf{62}, 054022 (2000).

\bibitem{Krelina:2018hmt}
   M.~Krelina, J.~Nemchik, R.~Pasechnik and J.~Cepila,
   ``Spin rotation effects in diffractive electroproduction of heavy quarkonia,''
   Eur. Phys. J. C \textbf{79}, 154 (2019).


\bibitem{Cepila:2019skb}
   J.~Cepila, J.~Nemchik, M.~Krelina and R.~Pasechnik,
   ``Theoretical uncertainties in exclusive electroproduction of S-wave heavy quarkonia,''
   Eur. Phys. J. C \textbf{79}, 495 (2019).


\bibitem{Krelina:2019egg}
   M.~Krelina, J.~Nemchik and R.~Pasechnik,
   ``$D$-wave effects in diffractive electroproduction of heavy quarkonia from the photon-like $V\rightarrow Q\bar Q$ transition,''
   Eur. Phys. J. C \textbf{80}, 92 (2020).

\bibitem{Krelina:2020bxt}
   M.~Krelina and J.~Nemchik,
   ``D-wave effects in heavy quarkonium production in ultraperipheral nuclear collisions,''
   Phys. Rev. D \textbf{102}, 114033 (2020).


\bibitem{Bertulani:2005ru}
   C.~A.~Bertulani, S.~R.~Klein and J.~Nystrand,
   ``Physics of ultra-peripheral nuclear collisions,''
   Ann. Rev. Nucl. Part. Sci. \textbf{55}, 271-310 (2005).


\bibitem{Nemchik:2002ug}
   J.~Nemchik,
   ``Incoherent production of charmonia off nuclei as a good tool for study of color transparency,''
   Phys. Rev. C \textbf{66}, 045204 (2002).

\bibitem{DeJager:1987qc}
   H.De Vries, C.W.De Jager and C.De Vries,
   ``Nuclear charge and magnetization density distribution parameters from elastic electron scattering,''
   Atom. Data Nucl. Data Tabl. \textbf{36}, 495 (1987).

\bibitem{Bauer:1977iq}
   T.~Bauer, R.~Spital, D.~Yennie and F.~Pipkin,
   ``The Hadronic Properties of the Photon in High-Energy Interactions,''
   Rev. Mod. Phys. \textbf{50}, 261 (1978).

\bibitem{Kopeliovich:1981pz}
   B.~Z.~Kopeliovich, L.~I.~Lapidus and A.~B.~Zamolodchikov,
   ``Dynamics of Color in Hadron Diffraction on Nuclei,''
   JETP Lett. \textbf{33}, 595 (1981).

\bibitem{Gribov:1968jf}
   V.~N.~Gribov,
   ``Glauber corrections and the interaction between high-energy hadrons and nuclei,''
   Sov.\ Phys.\ JETP {\bf 29}, 483 (1969)
   [Zh.\ Eksp.\ Teor.\ Fiz.\  {\bf 56}, 892 (1969)].

\bibitem{Kopeliovich:2016jjx}
   B.~Z.~Kopeliovich,
   ``Gribov inelastic shadowing in the dipole representation,''
   Int. J. Mod. Phys. A \textbf{31}, 1645021 (2016).

\bibitem{Hufner:1996dr}
   J.~Hufner, B.~Kopeliovich and J.~Nemchik,
   ``Glauber multiple scattering theory for the photoproduction of vector mesons off nuclei and the role of the coherence length,''
   Phys. Lett. B \textbf{383}, 362 (1996).


\bibitem{Kopeliovich:2005us}
   B.~Z.~Kopeliovich, I.~K.~Potashnikova and I.~Schmidt,
   ``Large rapidity gap processes in proton-nucleus collisions,''
   Phys. Rev. C \textbf{73}, 034901 (2006).

\bibitem{Bronzan:1974jh}
   J.~B.~Bronzan, G.~L.~Kane and U.~P.~Sukhatme,
   ``Obtaining Real Parts of Scattering Amplitudes Directly from Cross-Section Data Using Derivative Analyticity Relations,''
   Phys. Lett. B \textbf{49}, 272 (1974).

\bibitem{Forshaw:2003ki}
   J.~R.~Forshaw, R.~Sandapen and G.~Shaw,
   ``Color dipoles and rho, phi electroproduction,''
   Phys. Rev. D \textbf{69}, 094013 (2004).

\bibitem{Martin:1980jx}
   A.~Martin;
   ``A FIT of Upsilon and Charmonium Spectra,''
   Phys. Lett. B \textbf{93}, 338 (1980).

\bibitem{Barik:1980ai}
   N.~Barik and S.N.~Jena;
   ``Fine - Hyperfine Splittings Of Quarkonium Levels In An Effective Power Law Potential,''
   Phys. Lett. B \textbf{97}, 265 (1980). 

\bibitem{Buchmuller:1980su} 
   W.~Buchmuller and S.H.H.~Tye;
   ``Quarkonia and Quantum Chromodynamics,''
   Phys. Rev. D \textbf{24}, 132 (1981).  

\bibitem{GolecBiernat:1998js} 
   K.J.~Golec-Biernat and M.~Wusthoff,
   ``Saturation effects in deep inelastic scattering at low Q**2 and its implications on diffraction,''
   Phys. Rev. D \textbf{59}, 014017 (1998).
  
\bibitem{GolecBiernat:1999qd} 
   K.J.~Golec-Biernat and M.~Wusthoff,
   ``Saturation in diffractive deep inelastic scattering,''
   Phys. Rev. D \textbf{60}, 114023 (1999).

\bibitem{Bartels:2002cj}
   J.~Bartels, K.~J.~Golec-Biernat and H.~Kowalski,
   ``A modification of the saturation model: DGLAP evolution,''
   Phys. Rev. D \textbf{66}, 014001 (2002).

\bibitem{Ivanov:2002eq}
   Y.~P.~Ivanov, B.~Kopeliovich, A.~Tarasov and J.~Hufner,
   ``Electroproduction of charmonia off protons and nuclei,''
   AIP Conf. Proc. \textbf{660}, 283 (2003).


\bibitem{Melosh:1974cu} 
   H.J.~Melosh,
   ``Quarks: Currents and constituents,''
   Phys. Rev. D \textbf{9}, 1095 (1974).

\bibitem{Terentev:1976jk} 
   M.V.~Terentev,
   ``On the Structure of Wave Functions of Mesons as Bound States of Relativistic Quarks,''
   Sov.\ J.\ Nucl.\ Phys.\ \textbf{24}, 106 (1976)
   [Yad.\ Fiz.\ \textbf{24}, 207 (1976)].

\bibitem{Kopeliovich:2015qna} 
   B.Z.~Kopeliovich, E.~Levin, I.~Schmidt and M.~Siddikov,
   ``Lorentz-boosted description of a heavy quarkonium,''
   Phys. Rev. D \textbf{92}, 034023 (2015).

\bibitem{Hufner:1996jw} 
   J.~Hufner, B.~Kopeliovich and A.~B.~Zamolodchikov,
   ``Inelastic J / psi photoproduction off nuclei: Gluon enhancement or double color exchange?,''
   Z. Phys. A \textbf{357}, 113 (1997).

\bibitem{fg} 
   R.P.~Feynman and A.R.~Gibbs, 
   ``Quantum Mechanics and Path Integrals,'' 
   McGraw-Hill Book Company, NY 1965.


\bibitem{Kopeliovich:2001ee}
   B.~Kopeliovich, A.~Tarasov and J.~Hufner,
   ``Coherence phenomena in charmonium production off nuclei at the energies of RHIC and LHC,''
   Nucl. Phys. A \textbf{696}, 669 (2001).

\bibitem{Kopeliovich:2008ek}
   B.~Kopeliovich, J.~Nemchik, I.~Potashnikova and I.~Schmidt,
   ``Gluon Shadowing in DIS off Nuclei,''
   J. Phys. G \textbf{35}, 115010 (2008).

\bibitem{Krelina:2020ipn}
   M.~Krelina and J.~Nemchik,
   ``Nuclear shadowing in DIS at electron-ion colliders,''
   Eur. Phys. J. Plus \textbf{135}, 444 (2020).

\bibitem{guzey-1}
   V.~Guzey and M.~Zhalov,
   ``Exclusive $J/{\psi}$ production in ultraperipheral collisions at the LHC: constrains on the gluon distributions in the proton and nuclei,''
   JHEP \textbf{10}, 207 (2013).

\bibitem{guzey-2}
   V.~Guzey, E.~Kryshen, M.~Strikman and M.~Zhalov,
   ``Evidence for nuclear gluon shadowing from the ALICE measurements of PbPb ultraperipheral exclusive $J/{\psi}$ production,''
   Phys. Lett. B \textbf{726}, 290 (2013).


\bibitem{Nemchik:1997xb}
   J.~Nemchik, N.~N.~Nikolaev, E.~Predazzi, B.~G.~Zakharov and V.~R.~Zoller,
   ``The Diffraction cone for exclusive vector meson production in deep inelastic scattering,''
   J. Exp. Theor. Phys. \textbf{86}, 1054 (1998).

%
%


\bibitem{Afanasiev:2009hy}
   S.~Afanasiev \textit{et al.} [PHENIX],
   ``Photoproduction of J/psi and of high mass e+e- in ultra-peripheral Au+Au collisions at s**(1/2) = 200-GeV,''
   Phys. Lett. B \textbf{679}, 321 (2009).


\bibitem{Khachatryan:2016qhq}
   V.~Khachatryan \textit{et al.} [CMS],
   ``Coherent $J/\psi$ photoproduction in ultra-peripheral PbPb collisions at $\sqrt {s_{NN}} =$ 2.76 TeV with the CMS experiment,''
   Phys. Lett. B \textbf{772}, 489 (2017).
  
\bibitem{Abelev:2012ba}
   B.~Abelev \textit{et al.} [ALICE],
   ``Coherent $J/\psi$ photoproduction in ultra-peripheral Pb-Pb collisions at $\sqrt{s_{NN}} = 2.76$ TeV,''
   Phys. Lett. B \textbf{718}, 1273 (2013).
  

\bibitem{Abbas:2013oua}
   E.~Abbas \textit{et al.} [ALICE],
   ``Charmonium and $e^+e^-$ pair photoproduction at mid-rapidity in ultra-peripheral Pb-Pb collisions at $\sqrt{s_{\rm NN}}$=2.76 TeV,''
   Eur. Phys. J. C \textbf{73}, 2617 (2013).
  
  

\bibitem{Adam:2015sia}
   J.~Adam \textit{et al.} [ALICE],
   ``Coherent $\psi$(2S) photo-production in ultra-peripheral Pb Pb collisions at $\sqrt{s}_{\rm NN}$ = 2.76 TeV,''
   Phys. Lett. B \textbf{751}, 358 (2015).


\bibitem{Acharya:2019vlb}
   S.~Acharya \textit{et al.} [ALICE],
   ``Coherent J/$\psi$ photoproduction at forward rapidity in ultra-peripheral Pb-Pb collisions at $\sqrt{s_{\rm{NN}}}=5.02$ TeV,''
   Phys. Lett. B \textbf{798}, 134926 (2019).

\bibitem{ALICE:2021gpt}
   S.~Acharya \textit{et al.} [ALICE],
   ``Coherent $\rm{J/\psi}$ and $\rm{\psi'}$ photoproduction at midrapidity in ultra-peripheral Pb-Pb collisions at $\sqrt{s_{\mathrm{NN}}}~=~5.02$ TeV,''
   Eur. Phys. J. C \textbf{81}, 712 (2021).


\bibitem{LHCb:2018ofh}
   A.~Bursche [LHCb],
   ``Study of coherent $J/\psi$ production in lead-lead collisions at $\sqrt{s_{\rm NN}} =5\ \rm{TeV}$ with the LHCb experiment,''
   Nucl. Phys. A \textbf{982}, 247 (2019).


\bibitem{LHCb:2022ahs}
    R.~Aaij \textit{et al.} [LHCb],
    ``Study of coherent charmonium production in ultra-peripheral lead-lead collisions,''
    [arXiv:2206.08221 [hep-ex]].









\end{thebibliography}
\end{document}